\documentclass[pra,twocolumn,showpacs,preprintnumbers,floatfix]{revtex4-1}
\usepackage{graphicx}
\usepackage{dcolumn}
\usepackage{bm}
\usepackage{latexsym,epsfig}
\usepackage{graphicx}
\usepackage{verbatim}
\usepackage{comment}
\usepackage{amsmath}
\usepackage{amssymb}
\usepackage{stmaryrd}
\usepackage{color}
\usepackage{epstopdf}
\usepackage{float}
\usepackage[normalem]{ulem}
\DeclareGraphicsExtensions{.eps}

\begin{document}
\title{Three-body interacting dipolar bosons and the fate of lattice supersolidity}
\author{Manpreet Singh and Tapan Mishra}
\affiliation{Department of Physics, Indian Institute of Technology, Guwahati-781039, Assam, India}

\date{\today}

\begin{abstract}
We investigate a system of dipolar bosons in an optical lattice with local two and three-body interactions. Using the 
mean-field theory approach, we obtain the ground state phase diagram of the extended Bose-Hubbard (EBH) model with both 
repulsive and attractive three-body interactions. We show that the additional three-body on-site interaction has strong 
effects on the phase diagram especially on the supersolid phase. Positive values of the three-body interaction lead to 
the enhancement of the gapped phases at densities larger than unity by reducing the supersolid region. However, a small 
attractive three-body interaction enhances the supersolid phase.
\end{abstract}


\maketitle

\section{Introduction} 
The systems of ultracold quantum gases in optical lattices have been proven to be the most suitable platform for quantum 
simulation of complex many-body system due to the high degree of controllability of lattice parameters and geometry
~\cite{lewenstein_book}. Several interesting quantum phase transitions have been theoretically predicted and observed 
in the experiments in recent years. The seminal work on the quantum phase transition from the superfluid (SF) to the 
Mott insulator (MI) phase in dilute atomic gases~\cite{bloch} had paved the path to study numerous interesting quantum 
phenomena. The many-body physics which arises in such a dilute system is primarily attributed to the on-site two-body 
interaction.
However, recent developments show that multi-body effects can not be ruled out as they play very significant role on the 
ground state properties of the system~\cite{swill}. Based on the observations of collapse and revival of matter waves it 
has been shown that three-body interactions can play an important role in quantum phase transitions. 
Three- and higher-body interactions have been precisely measured for species like Rb~\cite{swill} and Cs~\cite{nagerl}
in different experiments. Recent studies show that it is also possible to engineer these multi-body interactions in a 
controlled way in systems of ultracold atoms~\cite{petrov1,petrov2,daley}. 
One can in principle have regimes where the the signs 
of the two-, three- and higher-body interactions can be selectively changed~\cite{petrov1,petrov2}.

On the other hand the long-range interaction in the polar lattice gases has been a topic of great interest in the past 
decade. Due to the non-local nature of the interaction, a wealth of new physics has been uncovered and one of them is the 
elusive supersolid (SS) phase of matter. The SS phase exhibits both superfluid and crystalline order which has attracted 
much attention since it was first proposed~\cite{andreev,chester}. The presence of non-local 
defects on top of a crystalline order gives rise to the superfluid order in a strongly interacting system~\cite{boninsegni}. 
It has also been predicted that supersolid may occur in lattices due to the long-range interaction. Several theoretical 
predictions have been made for various lattice models for the possible existence of the SS phase in the framework of 
the extended Bose-Hubbard (EBH) model~\cite{baranov2012rev}. Also, attempts have been made to look for a scenario 
where the lattice supersolidity can be achieved without considering
the non-local interactions in various conditions~\cite{aoki,huber,mishrasaw}.
Recent advances in the field of 
cooling and trapping of dipolar gases have opened up new directions to achieve longer range interaction 
in optical lattice~\cite{baranov2012rev,pfaureview}. Other possible candidates for such strong non-local interactions are the 
Rydberg dressed~\cite{pillet,pohl} atoms and polar molecules~\cite{jye}.

Recently the SS phase has 
been observed in $^{87}$Rb condensate coupled with optical cavity~\cite{esslinger1,esslinger2} where the long-range interaction 
is induced by the two photon processes.
 However, it's realisation in polar lattice gases 
is not yet achieved, which promises to be the most suitable system. 
Experimental progress in loading polar lattice gases in optical lattices~\cite{iskin7,koch} and most recently 
the realisation of the EBH model by loading dipolar gas of $^{168}$Er atoms in a three-dimensional 
optical lattice~\cite{ferlaino}, have paved the path to observe the exotic SS phase of matter. 

Although the SS phase can be achieved only in the presence of the on-site and nearest neighbour two body interactions, three-body interactions 
play very important role in deciding the fate of the supersolid phase. The effects of dominant two- and three-body 
local interactions in ultracold atomic systems have been studied extensively
~\cite{manpreet2,sowinski_3body,sowinski,zhang,scott1,greschner,souza,ejima,valencia1,kao}. It has also been shown that the 
cold polar molecules can interact via a long-range three-body 
interaction as well~\cite{zoller1,lauchli,Sansone_PRB79_020503,Wessel1,Tapan_PRA91_043614}, leading to interesting 
quantum phenomena. However, the effect of local three-body interaction on 
the phase diagram of the EBH model, especially on the supersolid phase has not been explored in detail. 

In this paper we consider a system of dipolar bosons with both on-site, two- and three-body interactions in an optical 
lattice, which can be explained by a modified EBH model. We analyse the ground state properties for both positive and 
negative three-body interactions with different combinations of the on-site and nearest neighbour repulsion. 
In our study, we show that the three-body interaction 
has very significant effect on the phase diagram of the EBH model. At incommensurate densities, repulsive three-body 
interaction enhances the gapped phases and simultaneously shrinks the SS region. However, the physics of the system becomes interesting 
when we consider a very small and attractive three-body interaction. In this case the supersolid phase becomes more robust. 
The SS region in the phase diagram gets enhanced by pushing the SS-SF boundary towards large hopping values. 
The three-body on-site interaction also introduces other gapped phases at commensurate densities due to the breaking of 
the degeneracies.

The remaining part of the paper in organised as follows. In Sec. II, we give details of the model and the method used in this work. 
Section III contains the discussion about the results and we conclude in Sec. IV.

\section{Model and method}
The effective many-body Hamiltonian which describes the system considered here is given by
\begin{eqnarray}
H=&-&t\sum_{<i,j>}(a_i^{\dagger}a_j+H.c.)+{\frac{U}{2}}\sum_{i}n_i(n_i-1)\nonumber\\
&+&V\sum_{<i,j>}n_i n_j+{\frac{W}{6}}\sum_i n_i(n_i-1)(n_i-2)\nonumber\\
&-&\mu\sum_{i}n_i
\label{eq:eq_one} 
\end{eqnarray}
where ${a_{i}}^{\dagger} (a_{i})$ are the bosonic creation (annihilation) operators at the site $i$, $t$ is the hopping 
amplitude between the adjacent sites $\langle i,j\rangle$, ${n}_i$ is the number operator at site $i$. $U$ and $W$ are 
the on-site two- and three-body interactions, $V$ is the nearest neighbour interaction (NNI) between the particles located at sites $i$ and 
$j$, and $\mu$ is the chemical potential. We investigate the ground state properties of the model (\ref{eq:eq_one}) using the 
single site mean-field decoupling approximation~\cite{sheshadri}. This method has been widely used to 
investigate the quantum phase transitions in systems of ultracold quantum gases in optical lattices. 
Although this method is more accurate for higher dimensions and large densities, where quantum fluctuations become less dominant, 
it gives enough information to qualitatively understand the underlying physics of the system in low dimensional cases. 
It is to be noted that in the case of dipolar atoms with repulsive $V$ in optical lattices, 
1D and 2D systems are more relevant from the experimental point of view. However, in order to understand an overall picture we 
investigate the global phase diagram for a d-dimensional lattice.

In the present case, using the MF decoupling approximation one can make the following substitutions,
\begin{eqnarray}
{a_{i}^{\dagger}a_{j}}&~&\simeq~\langle{a_{i}^{\dagger}}\rangle{a_j}+
{a_i}^{\dagger}\langle{a_j}\rangle-\langle{a_i}^{\dagger}\rangle\langle{a_j}\rangle\nonumber\\
{n_{i}^{\dagger}n_{j}}&~&\simeq~\langle{n_{i}^{\dagger}}\rangle{n_j}+
{n_i}^{\dagger}\langle{n_j}\rangle-\langle{n_i}^{\dagger}\rangle\langle{n_j}\rangle
\label{eq:eq_two}
\end{eqnarray}
in Eq.\ref{eq:eq_one} to write the single site MF Hamiltonian as:

\begin{eqnarray}
{{H_i^{MF}}\over{zt}}&\equiv&{{1}\over{2}} {{U}\over{zt}} n_i(n_i-1)-{\mu\over zt}n_i-(\phi_i a_i^\dag+\phi_i^* a_i)\nonumber\\
&+&{{1}\over{2}}(\psi_i^* \phi_i+\psi_i \phi_i^*)+ {{V}\over{t}}(n_i \overline{\rho}_i-\rho_i\overline{\rho}_i)\nonumber\\
&+&{{1}\over{6}}{{W}\over{zt}} n_i(n_i-1)(n_i-2),
\label{eq:eq_four}
\end{eqnarray}
where $\psi_{i}\equiv\langle{a_{i}^{\dagger}}\rangle\equiv\langle{a_i}\rangle$ and $\rho_{i}\equiv\langle{n_{i}}\rangle$ 
are the superfluid order parameter and density, respectively, $z(=2d)$ is the co-ordination number of a site in a $d$-dimensional 
hypercubic lattice, and 
$\phi_i\equiv{1 \over z} \sum_{i'} \psi_{i+i'}$, 
$\overline{\rho}_i\equiv{1 \over z} \sum_{i'} \rho_{i+i'}$; $i'$ takes into account the nearest-neighbours given by $z$.
Due to the possible presence of the DW order in the ground state, we consider a bipartite lattice configuration. 
We denote these sublattices by $\alpha$ and $\beta$ and write the MF Hamiltonian as the sum of the Hamiltonians
for these two sublattices.
\begin{eqnarray}
 H^{MF}=H^{MF}_\alpha+H^{MF}_\beta
 \label{eq:eq_five}
\end{eqnarray}
The $\phi$ and $\overline{\rho}$ terms in Eq.\ref{eq:eq_four} contain contribution from the neighbouring sites, that is, 
if $i=\alpha$, $\phi$ and $\overline{\rho}$ contain contribution from $\beta$-sublattice and vice versa. Therefore, we can 
write
\begin{eqnarray}
{{H_{\alpha}^{MF}}\over{zt}}&\equiv&{{1}\over{2}} {{U}\over{zt}} n_\alpha (n_\alpha-1)
-{\mu\over zt}n_\alpha-(\psi_\beta a_\alpha^\dag+\psi_\beta^* a_\alpha)\nonumber\\
&+&{{1}\over{2}}(\psi_\alpha^* \psi_\beta+\psi_\alpha \psi_\beta^*)+ {{V}\over{t}}(n_\alpha {\rho}_\beta-\rho_\alpha {\rho}_\beta)\nonumber\\
&+&{{1}\over{6}}{{W}\over{zt}} n_\alpha(n_\alpha-1)(n_\alpha-2),
\label{eq:eq_six}
\end{eqnarray}
Similarly, we obtain the expression for $H_{\beta}^{MF}$.
We then construct the Hamiltonian matrix for Eq.\ref{eq:eq_five} in the occupation number basis and diagonalize it
self-consistently to obtain the ground state energy and wavefunction. The ground-state wavefunction is then used to calculate 
the superfluid density ($\rho_s = \phi^2$) and number density ($\rho$) of the system. The maximum occupancy we
considered for our calculations is $9$ atoms per site. Hereafter we use bar($\textendash$) to denote the scaling by $zt$, 
e.g., $\overline{U}=U/zt$.

\section{Results and Discussion}
The phase diagram of the EBH model in the absence of $W$ is studied extensively in the recent years~\cite{baranov2012rev}. 
In the limit of negligible hopping and $z\overline V < \overline U$ the system exhibits alternate gapped DW 
and MI phases at commensurate densities. 
In the DW phases the alternate sites are occupied by the same number of particles due to the nearest-neighbour dipole-dipole 
repulsion. However, in the MI phase particle distribution is uniform throughout the lattice. 
On the other hand when $z\overline V > \overline U$, there exists only the DW phases. In both the limits, 
when the hopping amplitude $t$ increases, the SS 
phase starts to appear around the DW phases and further increasing the value of $t$ leads to the SF phase. 
Recent mean-field theory calculation using Gutzwiller approach predicts the above findings in the case of 
a d-dimensional hypercubic lattice~\cite{iskin}. It has also been shown that when the value of $\overline V$ increases, the SS region gets 
enlarged by pushing the SS-SF boundary. Similar studies in 1D, 2D and 3D systems have been carried out 
using powerful methods like the quantum Monte Carlo (QMC) and the density matrix renormalization group (DMRG) methods
~\cite{batrouni3,troyer3,kawashima,gangsu,tapan_pra043614}.

Using our mean-field calculation we first obtain the ground state phase 
diagram of the EBH model (with $\overline W=0$) for two regions of the NNI, one below the critical point 
$z\overline V<\overline U ~(z\overline V=0.9\overline U)$ and other above the critical point
$z\overline V>\overline U ~(z\overline V=1.5\overline U)$, as already discussed in Ref.~\cite{iskin}. Then we analyse the 
effect of both repulsive and attractive three-body interactions $\overline W$. Our main findings are depicted in 
Fig.~\ref{fig:fig1} and ~\ref{fig:fig2}. Before proceeding further we give the details of the abbreviations used for the 
gapped phases in the following table. 
\begin{center}
\begin{table}[!h]
\begin{tabular}{|l|c|r|}
\hline
Phase & Configuration & Density $\rho$ \\ \hline
MI1 & 1 1 1 1 1 & 1.0 \\ \hline
MI2 & 2 2 2 2 2 & 2.0 \\ \hline
DW1 & 1~0~1~0~1 & 0.5 \\ \hline
DW2 & 2~0~2~0~2 & 1.0 \\ \hline
DW3 & 2~1~2~1~2 & 1.5 \\ \hline
DW4 & 3~0~3~0~3 & 1.5 \\ \hline
DW5 & 3~1~3~1~3 & 2.0 \\ \hline
DW6 & 4~0~4~0~4 & 2.0 \\ \hline
DW7 & 3~2~3~2~3  & 2.5 \\ \hline
\end{tabular}
\caption{Density configuration of gapped MI and DW phases at commensurate densities.}
\label{table:tab1}
\end{table}
\end{center}

\begin{figure}[t]
  \centering
  \includegraphics[width=\columnwidth]{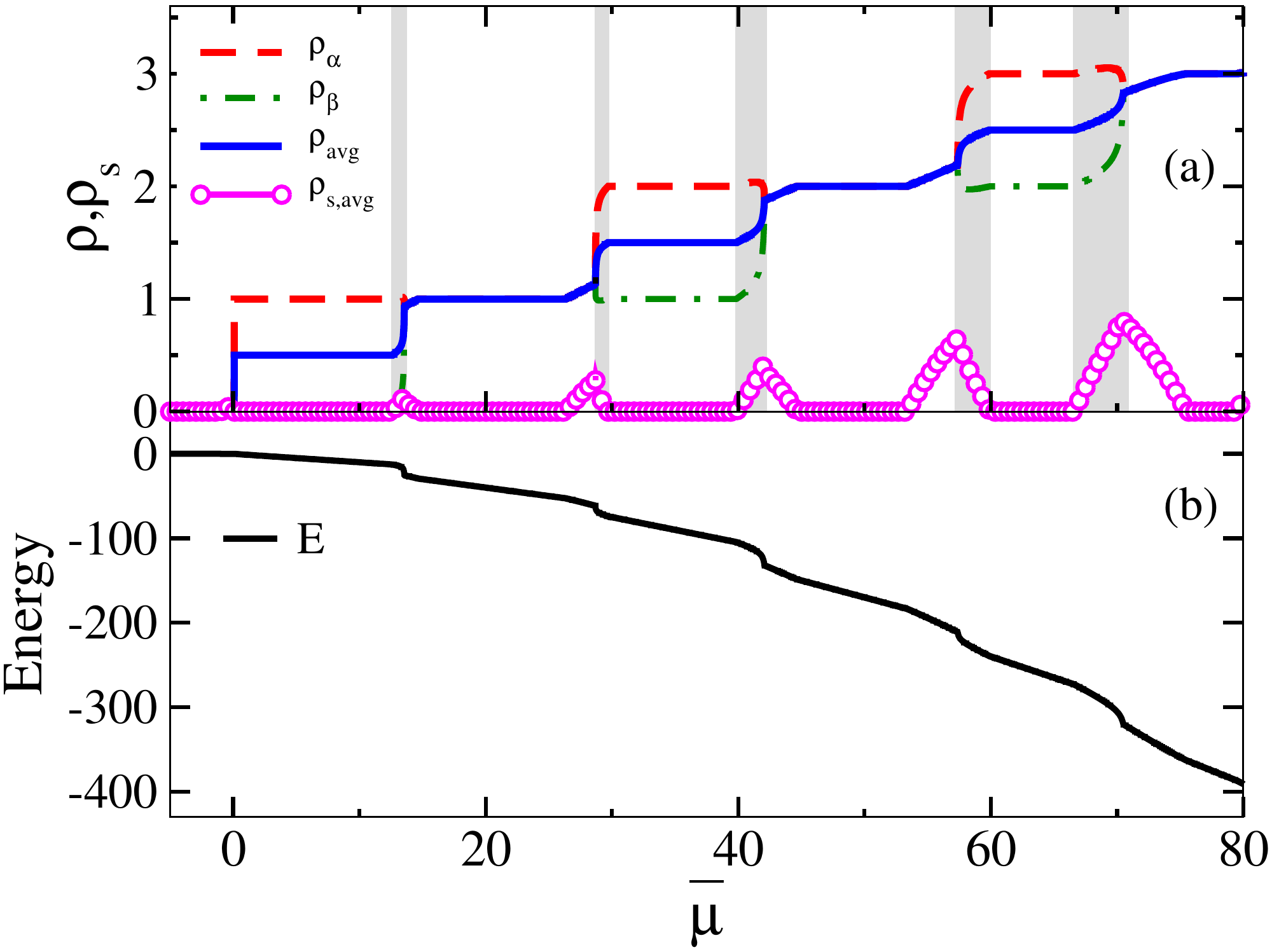}
    \caption{(Color online) (a)The sublattice densities $\rho_{\alpha ,\beta}$, the average density $\rho_{avg}$ and the average 
    superfluid density $\rho_{s,avg}$, as a function of the chemical potential $\overline\mu$ for $\overline U=15.0$, 
    $z\overline V=0.9\overline U$ and $\overline W=0$. The plateaus at commensurate densities show the gapped phases where 
    $\rho_s$ vanishes. The density imbalance on two sublattices shows the signature of the DW order. The SS phases are marked by
    grey shaded regions. 
    (b) The ground state energy plotted with respect to $\overline \mu$ showing a clear change in slope at the SF-SS transitions. 
    The slope changes at other phase transitions also (not clearly visible in the present scale of the axes).}
    \label{fig:rho}
\end{figure}
\begin{figure*}[htbp]
  \centering
  \includegraphics[width=\textwidth]{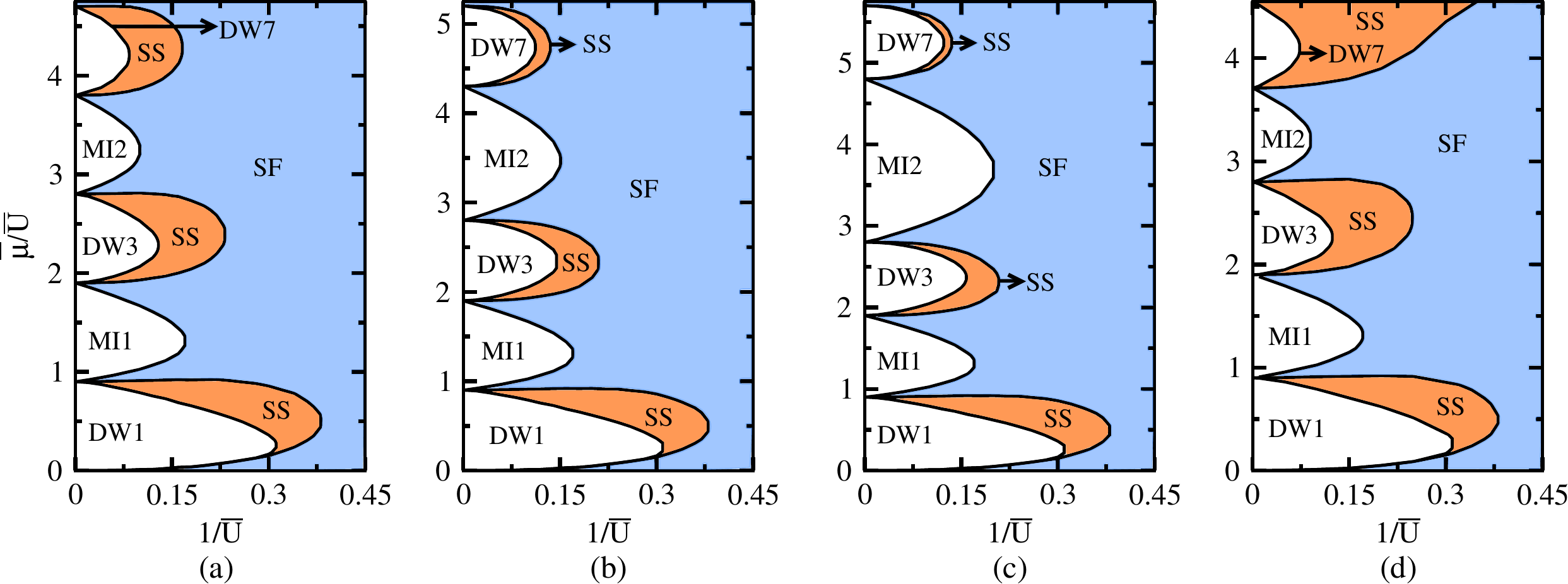}
    \caption{(Color online) Phase diagram of model~(\ref{eq:eq_one}) with repulsive and attractive three-body interaction $\overline W$ 
    for $z\overline V=0.9\overline U$. The values of $\overline W$ are (a)$\overline W=0$, 
    (b) $\overline W=0.5\overline U$, (c)$\overline W=1.0\overline U$ and (d) $\overline W=-0.09\overline U$.}
    \label{fig:fig1}
\end{figure*}
We use the average superfluid density $\rho_{s,avg} \big{(}=(\rho_{s,\alpha}+\rho_{s,\beta})\big{)}/2$ and average sublattice density 
$\rho_{avg} \big{(}=(\rho_{\alpha}+\rho_{\beta})\big{)}/2$ as a function of chemical potential to distinguish the gapped phases from the 
gapless phases. In the gapped phases $\rho_s$ vanishes and there appears a plateau in the chemical potential $\overline\mu$. 
The DW phases can be distinguished by the finite density imbalance between the $\alpha$ and $\beta$ sublattices. 
The SS phase is identified by looking for finite superfluid density and finite density imbalance as a function of the chemical 
potential. 

In the absence of the three-body interaction $\overline W$, as predicted before in Ref.~\cite{iskin}, for $z\overline V=0.9\overline U$ 
the system exhibits alternate gapped MI and DW phases at commensurate densities which is shown by plotting 
the densities with respect to the chemical potential in Fig.\ref{fig:rho}(a). We plot $\rho_\alpha$ (red dashed line), 
$\rho_{\beta}$ (green dot-dashed line), $\rho_{avg}$ (blue solid line) and 
$\rho_{s,avg}$ (magenta circles) 
with respect to $\overline\mu$ for $\overline{U}=15.0$ and $z\overline V=0.9\overline U$. It can be seen that 
the plateaus appear at $\rho_{avg}=0.5, 1.0, 1.5, 2.0, 2.5$ corresponding to the
gapped phases DW1, MI1, DW3, MI2 and DW7 phases, respectively. The associated  
arrangement of atoms in sublattices $\alpha$ and $\beta$ are shown in the Table \ref{table:tab1}. In the plateau regions the superfluid density 
$\rho_{s,avg}$ vanishes. 
The DW phases are identified by looking at the sublattice densities $\rho_{\alpha}$ and $\rho_{\beta}$.
The region around the plateaus are gapless SF phases. However, there are certain regions where there exists 
finite sublattice density imbalance in the 
SF regions. These regions correspond to the SS phase as discussed before. 

By repeating the calculation for different values of $\overline U$ we obtain the complete phase diagram of the model given in Eq.(\ref{eq:eq_one}) 
for $\overline W=0$ (Fig.\ref{fig:fig1}(a), previously shown in Ref.~\cite{iskin}) and two different values of $\overline W=0.5\overline U$,
$1.0\overline U$ in Fig.\ref{fig:fig1}(b), (c), respectively. The phase diagrams are obtained by locating the end points of the plateaus for the gapped 
phases and for the supersolid phase we look for the points where the density imbalance disappears inside the SF phase. 
In Fig.\ref{fig:fig1}(a), for $\overline W=0$, the gapped MI and DW lobes are formed in the strong interaction regime. In the weak interaction
regime the SF phase appears and the DW lobes are surrounded by the SS phases.  
In our analysis we truncate the phase diagram at the DW lobe of density $2.5$ as it is sufficient for the relevant physics.  

\subsection{Repulsive 3-body interaction}
By turning on the three-body interaction $\overline W$ we see a clear effect on the quantum phases at large densities as shown in the 
phase diagrams. In Fig.\ref{fig:fig1}(b) and (c) it can be seen that the gapped DW3, MI2 and DW7 lobes become larger and the SS phases shrink 
drastically as $\overline W$ increases. This can be understood
as follows: although the range of $\overline U$ and $\overline V$ is fixed, the positive values of $\overline W$ increase the effective 
onsite interaction. As a result the gapped phases lobes, such as that of the MI and DW, increase in size and the SS region shrinks.
We repeat the same calculation but with $z\overline V=1.5\overline U$ and show the result in Fig.\ref{fig:fig2}(a)-(c). 
In the absence of $\overline W$, as predicted in Ref.~\cite{iskin}, 
the MI phases disappear and only the DW phases such as DW1, DW2, DW4 and DW6 appear as shown in Fig.\ref{fig:fig2}(a). 
The SS phase in this case occupies a 
large region of the phase diagram. The effect of repulsive three-body interaction can be seen in the phase diagrams given in 
Fig.\ref{fig:fig2}(b) and Fig.\ref{fig:fig2}(c).
For $\overline W=0.5\overline U$ the size of DW2 and  DW4 lobes increases, whereas the DW6 phase 
disappears due to the effect of the repulsive $W$ as shown 
in Fig.\ref{fig:fig2}(b). 
At the same time DW5 phase appears in the phase diagram. 
The SS phase shrinks drastically similar to the case for $z\overline V < \overline U$. 
The SS-SF boundary in this case seems 
to close after the DW5 lobe unlike the case for $\overline W=0$ where the 
SS-SF boundary grows almost linearly. When $\overline W$ is increased further i.e. for $\overline W=1.0\overline U$ the on-site
repulsion becomes more dominant and hence the system does not allow individual lattice sites to have more than two atoms. This 
leads to the disappearance of the DW4 phase and appearance of the DW3 phase. Surprisingly, the competition between $\overline W$
and $\overline V$ leads to the appearance of MI2 phase which is absent with only local two-body interaction. The SS phase 
shrinks and envelopes all the DW phases as shown in Fig.\ref{fig:fig2}(c). 
The SF-DW phase transition below $\rho =0.5$ is a first order transition, where as the SS-SF transition and DW-SS transitions are 
found to be continuous~\cite{ssinha}. The signatures of these phase transitions can be 
seen from the $\rho$ vs. $\mu$ plot as shown in 
Fig.\ref{fig:rho}(a). The corresponding ground state energy with respect to $\overline \mu$  also shows the change in 
slope at different phase transitions points as shown in Fig.\ref{fig:rho}(b). 

\subsection{Attractive 3-body interaction}
In this section we 
discuss the effect of attractive three-body interaction. It is well known that a small attractive local interaction leads to 
the collapse of atoms onto a single lattice site~\cite{dalfovormp}. However, due to the presence of two-body repulsion, the 
system does not collapse up to a certain threshold value of $\overline W$. For the situation considered here, 
i.e. for $z\overline V=0.9\overline U$ the collapse occurs at $\overline W=-0.21\overline U$. 
\begin{figure*}[htbp]
  \centering
  \includegraphics[width=\linewidth]{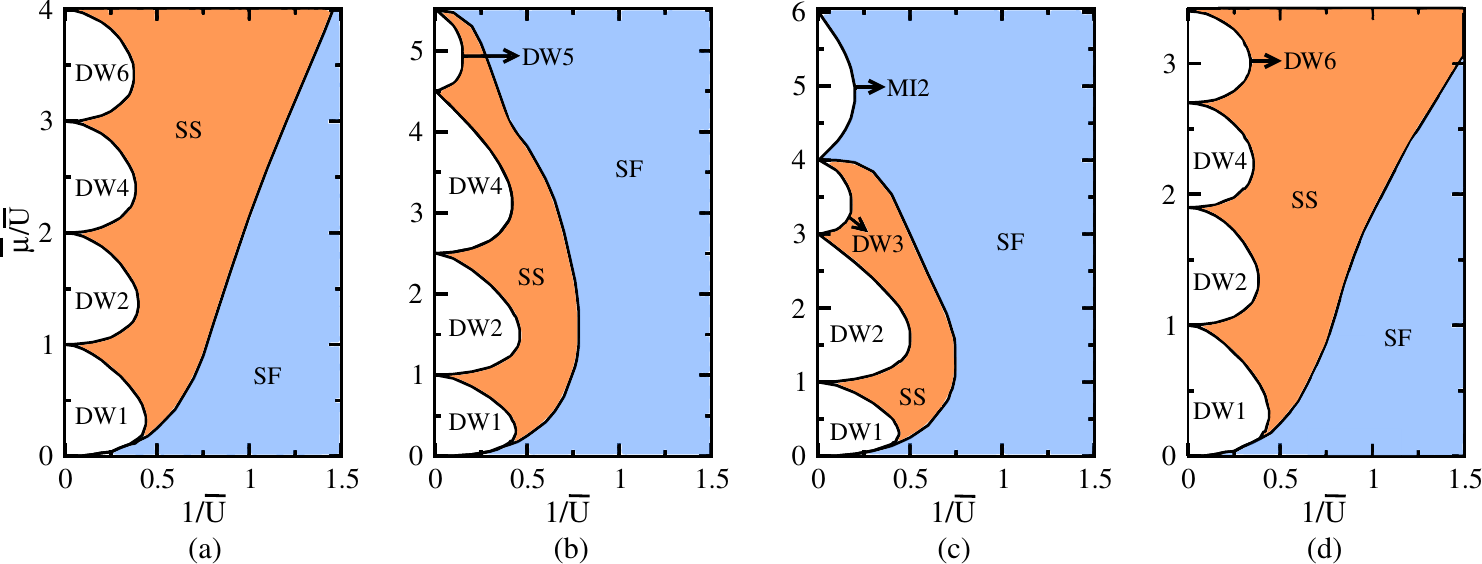}
    \caption{(Color online) Phase diagram of model~(\ref{eq:eq_one}) with repulsive and attractive three-body interaction $\overline W$ 
    for $z\overline V=1.5\overline U$. The values of $\overline W$ are (a)$\overline W=0$, (b) $\overline W=0.5\overline U$, (c)$\overline W=1.0\overline U$ and 
    (d) $\overline W=-0.1\overline U$. }
    \label{fig:fig2}
\end{figure*}
Hence, in this case we restrict ourselves to very small negative values of $\overline W$ compared to the repulsive two-body 
interaction $\overline U$. We find that though $\overline W$ is very small, it has a 
prominent effect on the phase diagram. Interestingly, the attractive three-body interaction favours the SS phase which can 
be seen from Fig.~\ref{fig:fig1}(d) and Fig.~\ref{fig:fig2}(d) for $z\overline V=0.9\overline U, \overline W=-0.09\overline U$ 
and $z\overline V=1.5\overline U, \overline W=-0.1\overline U$ respectively.
It can be seen that when $z\overline V<\overline U$ ($z\overline V=0.9\overline U$), there is a small reduction in the DW3 and 
DW7 lobes (Fig.~\ref{fig:fig1}(d)) as compared to the case of $\overline W=0$ (Fig.~\ref{fig:fig1}(a)). 
The tip of the DW3 phase shifts from $1/\overline U=0.13$ to $0.12$ where as the tip of DW7 phase shifts from $1/\overline U=0.84$ to $0.73$.
This reduction is obvious as the attractive three-body interaction reduces the effective on-site interaction by a small amount. This 
in turn reduces the gapped phases at higher densities as predicted in Ref.~\cite{sowinski_3body} for the Bose-Hubbard model. 

However, the competition between the strong repulsive $\overline U$, $\overline V$ and weak attractive $\overline W$ modifies 
the SS phase drastically. The SS regions enlarge in the vicinity of the DW phases and the SS-SF boundary shifts towards the 
weak coupling regimes as shown in Fig.~\ref{fig:fig1}(d). By comparing Fig.~\ref{fig:fig1}(a) and (d), 
it can be seen that the tip of the SS region (surrounding the DW3 lobe) 
shifts from $1/\overline U\approx0.23$ to $0.25$. 
The width of SS region also increases, e.g. at $1/\overline U=0.2$, $(\mu/U)_{min}$ changes from $2.1$ to $2.09$ and
$(\mu/U)_{max}$ changes from $2.68$ to $2.78$. Here $(\overline \mu/\overline U)_{min}$ and 
$(\overline \mu/\overline U)_{max}$ are the points on the lower and upper
boundary of the SS region, respectively. This effect is more prominent for larger densities as seen in the 
region around the DW7 phase as shown in Fig.~\ref{fig:fig1}(d). 
The enhancement of the SS phase is very similar to the case when $z\overline V > \overline U$ and 
$\overline W =0$ as predicted in Ref.~\cite{iskin}. This finding shows that a small negative three-body interaction can actually 
play a very important role in enlarging the SS phase even without making the dipole-dipole interaction $z\overline V$ stronger 
than the on-site interaction $\overline U$ at large densities, which is a relevant experimental condition. For the case of 
$z\overline V = 1.5 \overline U$ and $\overline W=-0.1\overline U$, similar effects are seen. The DW phases at higher densities 
such as the DW2, DW4 and DW6 phases shrink and there is a clear enhancement of the SS phase in the vicinity of the DW4 and DW6 
phases which can be seen by comparing Fig.\ref{fig:fig2}(a) and Fig.\ref{fig:fig2}(d). For example, at 
$1/\overline{U}=1.0$, $(\overline \mu/\overline U)_{min}$ changes from $2.15$ to $1.85$ and at $1/\overline{U}=1.5$ it 
changes from $4.18$ to $3.06$.
The enhancement of the SS phase can be attributed to reduction in the effective on-site repulsion due to the attractive three-body interaction 
while the nearest neighbour interaction remains unaffected. As a result the defects in the system become more mobile on top of the 
DW phase and hence the supersolid nature increases. Therefore, the SS-SF phase boundary shifts towards the large values of 
$1/\overline U$. It is to be noted that the system gradually collapses by making $\overline W$ further attractive.

\section{Conclusions}
We investigate a system of dipolar bosons in a $d$-dimensional lattice in the framework of the modified EBH model and study the 
effect of the three-body local interaction on the supersolid phase using the self consistent 
mean-field decoupling approximation. Considering two limiting cases such as $z\overline V=0.9\overline U$ and $z\overline V=1.5\overline U$, 
we show that the phase diagram changes drastically 
due to the effect of both repulsive and attractive three-body interactions. In the case of repulsive $\overline W$, the lobes corresponding to the gapped
phases such as the MI and the DW phase enlarge at densities larger than unity. As a result the SS phase shrinks substantially. 
When $z\overline V>\overline U$, new DW phases appear due to the breaking of the degeneracies. Also the MI phase reappears when 
$\overline W=1.0\overline U$ which was not the case when $\overline W=0$. In this case the SS phase tends to 
shrink and surround multiple DW lobes in contrast to the case of $\overline W=0$ where the SS phase expands pushing the SS-SF boundary almost linearly~\cite{iskin}.
However, the situation is 
completely different when a very small attractive three-body interaction is introduced. Interestingly, the small attractive $\overline W$ 
favours the SS phase which enlarges further as compared to the $\overline W=0$ case.

As discussed before, the effects of three-body on-site interaction in ultracold atomic systems are generally 
considered to be small. 
However, recent progress in theoretical studies 
show that the three-body interaction can be engineered under proper conditions~\cite{tiesinga,petrov1,petrov2,daley}. 
Three-body interaction can also be made negative keeping the
two-body interaction positive~\cite{tiesinga,petrov1,petrov2}. In principle, it is possible to tune the two and three-body interactions 
independently with respect to the off-site dipole-dipole interaction. Hence, our findings provide further scopes to observe the effect of the local three-body 
interaction on the phase diagram of the EBH model and in particular the exotic supersolid phase 
of matter using tunable attractive three-body interaction at large densities. 

\section{Acknowledgement}
We would like to thank Luis Santos, Paolo Pedri, Subroto Mukerjee, Ramesh V. Pai, Sebastian Greschner and Sebastian Will 
for the useful discussions.
MS would like to acknowledge support from DST-SERB, India for the financial support through project number PDF/2016/000569. 
TM acknowledges the support by the start-up research grant from the Indian Institute of Technology, Guwahati,
India. Computational work was done using the HPC (Param Ishan) facility at the Indian Institute of Technology, Guwahati.

\end{document}